# Autonomous Vehicle Convoy Control as a Differential Game

Hossein B. Jond, and Jan Platos, *Member, IEEE,*

*Abstract*—Group control of connected and autonomous vehicles on automated highways is challenging for the advanced driver assistance systems (ADAS) and the automated driving systems (ADS). This paper investigates the differential game-based approach to autonomous convoy control with the aim of deployment on automated highways. Under the non-cooperative differential games, the coupled vehicles make their decisions independently while their states are interdependent. The receding horizon Nash equilibrium of the linear-quadratic differential game provides the convoy a distributed state-feedback control strategy. This approach suffers a fundamental issue that neither a Nash equilibrium's existence nor the uniqueness is guaranteed. We convert the individual dynamics-based differential game to a relative dynamics-based optimal control problem that carries all the features of the differential game. The existence of a unique Nash control under the differential game corresponds to a unique solution to the optimal control problem. The latter is shown, as well as the asymptotic stability of the closed-loop system. Simulations illustrate the effectiveness of the presented convoy control scheme and how it well suits automated highway driving scenarios.

## I. INTRODUCTION

ON automated highway driving, collective traveling of autonomous vehicles enhances road capacity, reduces fuel consumption, increases safety, and improves user comfort [1]. Platoon and convoy are the two common forms of group driving in which vehicles reach a consensual speed at the motion direction while maintaining priorly specified distances to their adjacents. The former refers to the longitudinal control of a group of autonomous vehicles moving in the same lane of the highway [2]. The latter refers to the longitudinal-lateral control of vehicles over different lanes and is more challenging. Autonomous vehicles can form a convoy by acquiring and maintaining a formation. The control algorithm adjusts the individual vehicles' control inputs to ensure they achieve (and keep) desired formation shape in lateral and longitudinal directions. Coupled vehicles control algorithms demand information exchange between the vehicles through inter-vehicle communications such as their positions and velocities. Most of the convoy control ideas come from multi-agent formation control algorithms [3]. Besides the group lateral maneuvers in automated highways, convoys may have other applications such as protection of VIP vehicles and snow plowing [4].

A vehicle has complex dynamics and kinematics. Some select a dynamics model with fewer details that approximate most of a vehicle's behavior [5]. Simplified dynamics models such as bicycle, single- and double-integrator models are widely used [6]–[8]. A vehicle with double-integrator dynamics is a two-dimensional linear point mass with acceleration as the control input [9]. Autonomous driving systems require a feedback controller to react instantaneously to the current situation. On automated highway driving, a vehicle sense the highway and other vehicles through the onboard sensors, Vehicle-to-Vehicle (V2V), or Vehicle-to-Infrastructure (V2I) communications [10], [11]. Although V2V/V2I communications are not vital to convoy control, receiving the position and velocity data via those channels can reduce the errors from the onboard sensors. The inter-vehicle communications of a convoy can be modeled using graph theory. The data from V2V/V2I communications and onboard sensors usually are collected and transferred in the receding horizon manner [12]. In this control scheme, during each update time, each vehicle receives and transmits information with the immediate neighbors (and the infrastructure) then minimizes its control by solving the assigned open-loop control problem [13].

Differential games are prominent tools for behavior analysis of a team of self-interested and competitive dynamical agents/players [14]. Each player solves its optimal control problem and shares its control strategy with some (or all) of the players. Therefore, the evolution of the states of the players is impacted by each other. Game theory provides an equilibrium solution that is in the best interest of all players. The linear-quadratic differential games are popular due to their analytic tractability and the possibility to calculate a linear control law [15]. Multi-agent control problems such as formation [16], consensus [17], [18], and synchronization [19] already have been investigated using these games. An autonomous vehicle is a self-interested decision-maker. It can decide on its control strategy based on personal interests. In terms of convoy control, each vehicle's interest could be to penalize its relative displacement and velocity errors taking its interests such as the fuel amount into account [20].

The use of the linear-quadratic Nash differential game to model formation control first has been reported in [21]. The report shows that the Nash equilibrium for each robot is a distributed control as it requires information only from the immediate neighbors on the communication topology. The existence and uniqueness of the Nash equilibria in differential/dynamic games are not ensured at the outset [22]. A set of coupled (partial) differential matrix equations must be solved, which is not generally straightforward [23].

In this paper, an autonomous convoy with self-interested vehicles is formulated as a non-cooperative linear-quadratic differential game. The Nash equilibrium of the game is a

H.B. Jond and J. Platos are with the Department of Computer Science, VSB-Technical University of Ostrava, Ostrava, Czech Republic, e-mail: {hossein.barghi.jond, jan.platos}@vsb.cz.

mutually beneficial control strategy, and nobody will deviate from it. Therefore, it can be exploited as a self-enforcing convoy control strategy. The main contributions of the paper in comparison with the similar works in [21], [24], [25] are as follows:

1) Presuming it, [21] does not prove the existence and uniqueness of the Nash equilibrium. We prove the existence and uniqueness of the Nash equilibrium by converting the game problem to an optimal control problem via relative dynamics transformation.
2) [24] investigates the existence and uniqueness of the Nash equilibrium for simplified quadratic costs, so the problem is not analytically intricate.
3) [25] investigates the existence and uniqueness of the Nash equilibrium by replacing the continuous-time differential game with a discrete-time dynamic game. We tackle the continuous-time game problem that is highly analytically intricate.
4) [24], [25] discuss the open-loop control strategy while this work develops a state-feedback receding horizon Nash controller for practical situations.
5) [21], [24], [25] focus on multi-robot formation control while this work promotes the differential game usages to group control of self-interested autonomous vehicles for automated highway driving scenarios.

The rest of this paper is organized as follows. Section II provides preliminaries. The differential game model of convoy control is introduced in Section III. In Section IV, the differential game is converted to an optimal control problem via defining relative dynamics and, the analytical results on the existence and uniqueness of the solution to the optimal control problem are presented. The state-feedback system and the asymptotic stability of the receding horizon-based closed-loop system are discussed in Section V. Section VI illustrates the simulation results for automated highway driving scenarios with and without state constraints. The conclusion is given in Section VII.

## II. Preliminaries

### A. Graph Theory

A directed graph $\mathcal{G}(\mathcal{V}, \mathcal{E})$ consists of a set of vertices $\mathcal{V} = \{1, 2, \ldots, m\}$ and edges $\mathcal{E} \subseteq \{(i,j) : i,j \in \mathcal{V}, j \neq i\}$ (i.e., ordered pairs of distinct vertices and has no self-loops). Each edge $(i,j) \in \mathcal{E}$ is assigned with a positive scalar weight $\mu_{ij}$. The set of neighbors of vertex $i$ is defined by $\mathcal{N}_i = \{j \in \mathcal{V} : (i,j) \in \mathcal{E}\}$. Graph $\mathcal{G}$ is connected if for every pair of vertices $(i,j) \in \mathcal{V}$, from $i$ to $j$ for all $j \in 1, \ldots, m, j \neq i$, there exists a path of (undirected) edges from $\mathcal{E}$.

Matrix $\boldsymbol{D} \in \mathbb{R}^{m \times |\mathcal{E}|}$ is the incidence matrix of $\mathcal{G}$ where $\boldsymbol{D}$'s $uv$th element is 1 if the node $u$ is the head of the edge $v$, -1 if the node $u$ is the tail, and 0, otherwise. The Laplacian matrix is defined as:

$$\boldsymbol{L} = \boldsymbol{D}\boldsymbol{W}\boldsymbol{D}^T \quad (1)$$

where $\boldsymbol{W} = \text{diag}(\ldots, \mu_{ij}, \ldots) \in \mathbb{R}^{|\mathcal{E}|}, \forall (i,j) \in \mathcal{E}$.

The graph Laplacian $\boldsymbol{L} \in \mathbb{R}^m$ is symmetric ($\boldsymbol{L} = \boldsymbol{L}^T$), positive semidefinite ($\boldsymbol{L} \geq 0$), and satisfies the sum-of-squares property [26]:

$$\sum_{(i,j) \in \mathcal{E}} \mu_{ij} \|x_i - x_j\|^2 = \boldsymbol{x}^T \boldsymbol{L} \boldsymbol{x} \quad (2)$$

where $\boldsymbol{x} = [x_1^T, \ldots, x_m^T]^T \in \mathbb{R}^m$ and $\|.\|$ is the Euclidean norm. The Kronecker product is used to extend the dimension of the Laplacian.

### B. Kronecker Product and Sum

Let $\boldsymbol{X} \in \mathbb{R}^{m \times n}, \boldsymbol{Y} \in \mathbb{R}^{p \times q}$. The Kronecker product is defined as:

$$\boldsymbol{X} \otimes \boldsymbol{Y} = \begin{bmatrix} x_{11}\boldsymbol{Y} & \cdots & x_{1n}\boldsymbol{Y} \\ \vdots & \ddots & \vdots \\ x_{m1}\boldsymbol{Y} & \cdots & x_{mn}\boldsymbol{Y} \end{bmatrix} \in \mathbb{R}^{mp \times nq} \quad (3)$$

The Kronecker product has the following properties [31]:

$$(\boldsymbol{X} \otimes \boldsymbol{Y})^T = \boldsymbol{X}^T \otimes \boldsymbol{Y}^T \quad (4)$$

$$(\boldsymbol{X} \otimes \boldsymbol{Y})^{-1} = \boldsymbol{X}^{-1} \otimes \boldsymbol{Y}^{-1} \quad (5)$$

$$(\boldsymbol{X} \otimes \boldsymbol{Y})(\boldsymbol{U} \otimes \boldsymbol{V}) = \boldsymbol{X}\boldsymbol{U} \otimes \boldsymbol{Y}\boldsymbol{V} \quad (6)$$

$$\det(\boldsymbol{X} \otimes \boldsymbol{Y}) = (\det \boldsymbol{X})^m (\det \boldsymbol{Y})^n, \quad \boldsymbol{X} \in \mathbb{R}^n, \boldsymbol{Y} \in \mathbb{R}^m \quad (7)$$

$$e^{\boldsymbol{X} \otimes \boldsymbol{Y}} = e^{\boldsymbol{Y}} \otimes e^{\boldsymbol{X}} \quad (8)$$

Let $\boldsymbol{X} \in \mathbb{R}^n, \boldsymbol{Y} \in \mathbb{R}^m$. The Kronecker sum is defined as:

$$\boldsymbol{X} \oplus \boldsymbol{Y} = (\boldsymbol{I}_m \otimes \boldsymbol{X}) + (\boldsymbol{Y} \otimes \boldsymbol{I}_n) \in \mathbb{R}^{mn} \quad (9)$$

## III. Convoy Control as a Differential Game

A group of vehicles with double integrator dynamics and quadratic costs of acquiring and maintaining a convoy can be formulated as a linear-quadratic Nash differential game. Then, the main issue that arises here is whether a unique Nash equilibrium exists which we will cover in the next section. The convoy control problem definition here is from [21], [24], [25].

Consider a networked convoy of $m$ vehicles, each of which modeled by a double integrator dynamics:

$$\ddot{\boldsymbol{q}}_i = \boldsymbol{u}_i \quad (10)$$

where $\boldsymbol{q}_i, \boldsymbol{u}_i \in \mathbb{R}^2$ denote the longitudinal-lateral position and control input vector of vehicle $i \in \{1, \ldots, m\}$, respectively. Here, the positions and velocities are the system states and the accelerations are the controls.

Let $\boldsymbol{z} = [\boldsymbol{q}_1^T, \ldots, \boldsymbol{q}_m^T, 1, \dot{\boldsymbol{q}}_1^T, \ldots, \dot{\boldsymbol{q}}_m^T]^T \in \mathbb{R}^{4m+1}$ be the convoy state vector. The convoy motion can be expressed by system dynamics,

$$\dot{\boldsymbol{z}} = \boldsymbol{A}\boldsymbol{z} + \sum_{i=1}^m \boldsymbol{B}_i \boldsymbol{u}_i \quad (11)$$

where $\boldsymbol{A} = \begin{bmatrix} \boldsymbol{o} & \boldsymbol{I}_{2m} \\ \boldsymbol{o} & \boldsymbol{o} \end{bmatrix} \in \mathbb{R}^{4m+1}, \boldsymbol{B}_i = [\boldsymbol{o}_{2m+1 \times 2}, \boldsymbol{b}_i]^T \in \mathbb{R}^{4m+1 \times 2}, \boldsymbol{b}_i = [\boldsymbol{o}_2, \ldots, \boldsymbol{I}_2, \ldots \boldsymbol{o}_2]^T \in \mathbb{R}^{2m \times 2}, \boldsymbol{I}_m \in \mathbb{R}^m$ and $\boldsymbol{o}_m \in \mathbb{R}^m$ are the identity matrix and zero vector/matrix of dimension $m$.

The information flow among the members is necessary to convoy control. The communication network is represented by the convoy graph $\mathcal{G}(\mathcal{V}, \mathcal{E})$.

**Assumption 1.** *The communication network graph $\mathcal{G}$ is connected.*

Vehicle $i \in \{1, \ldots, m\}$ can acquire and keep the desired relative position within the convoy by driving only the relative displacements and relative velocities with respect to its neighbor vehicle $j \in \mathcal{N}_i$ to the desired distance $d_{ij} \in \mathbb{R}^2$ and zero, respectively. The local error for vehicle $i$ to be minimized is defined as [24], [25]:

$$\sum_{j \in \mathcal{N}_i} \mu_{ij}(\|q_i - q_j - d_{ij}\|^2 + \|\dot{q}_i - \dot{q}_j\|^2) = z^T Q_i z \quad (12)$$

where
$$Q_i = \begin{bmatrix} L_i \otimes I_2 & -(DW_i \otimes I_2)d & o \\ -((DW_i \otimes I_2)d)^T & d^T(W_i \otimes I_2)d & o \\ o & o & L_i \otimes I_2 \end{bmatrix},$$
$L_i = DW_iD^T$, $W_i = \mathrm{diag}(0, \ldots, \mu_{ij}, \ldots, 0) \in \mathbb{R}^{|\mathcal{E}|}$, $j \in \mathcal{N}_i$ and $d = [\ldots, d_{ij}^T, \ldots]^T \in \mathbb{R}^{2|\mathcal{E}|}$, $(i,j) \in \mathcal{E}$. The matrix form of the error is obtained using the sum-of-squares property of Laplacian in (2).

Over a finite-planning horizon $t_f$, vehicle $i$ can acquire and maintain its desired state within the convoy by minimizing the following quadratic cost [24], [25]:

$$J_i = z^T(t_f) Q_{if} z(t_f) + \int_0^{t_f} (z^T Q_i z + \sum_{j=1}^{m} u_j^T R_{ij} u_j) dt \quad (13)$$

where
$$Q_{if} = \begin{bmatrix} L_{if} \otimes I_2 & -(DW_{if} \otimes I_2)d & o \\ -((DW_{if} \otimes I_2)d)^T & d^T(W_{if} \otimes I_2)d & o \\ o & o & L_{if} \otimes I_2 \end{bmatrix}$$
, $L_{if} = DW_{if}D^T$, $W_{if} = \mathrm{diag}(0, \ldots, \omega_{ij}, \ldots, 0) \in \mathbb{R}^{|\mathcal{E}|}$, $j \in \mathcal{N}_i$ and $\omega_{ij} > 0$. We assume that $R_{ii}$ is positive definite ($R_{ii} > 0$). Therefore, the control objective for vehicle $i$ is to design $u_i$ to minimize its cost function $J_i$ for the underlying individual dynamics system (11).

The cost function $J_i$ is shared among the other vehicles. As $J_i \neq J_k$ for $i, k \in 1, \ldots, m$ and $i \neq k$, there are m different cost functions, the problem poses as a differential game. Here, the linear dynamical system (11) and quadratic cost function (13) can be viewed as the state dynamics and objective function of player $i$ of a non-cooperative linear-quadratic differential Nash game [21]. A Nash equilibrium is a strategy combination of all players in the game with the property that no one can gain lower cost by unilaterally deviating from it. The open-loop Nash equilibrium is defined as a set of admissible actions $(u_1^*, \ldots, u_m^*)$ if for all admissible $(u_1, \ldots, u_m)$ the following inequalities $J_i(u_1^*, \ldots, u_{i-1}^*, u_i^*, u_{i+1}^*, \ldots, u_m^*) \leq J_i(u_1^*, \ldots, u_{i-1}^*, u_i, u_{i+1}^*, \ldots, u_m^*)$ hold for $i = 1, \ldots, m$ where $u_i \in \Gamma_i$ and $\Gamma_i$ is the admissible strategy set for player $i$.

The definition of non-cooperative linear-quadratic differential games and their solvability conditions for a unique Nash equilibrium can be found in [23], chapter 7.

Define: $S_i = B_i R_{ii}^{-1} B_i^T$, $M = \begin{bmatrix} -A & S_1 & \cdots & S_n \\ Q_1 & A^T & o & o \\ \vdots & & \ddots & o \\ Q_m & o & o & A^T \end{bmatrix}$

and $H(t_f) = \begin{bmatrix} I_{4m} & o & \cdots & o \end{bmatrix} e^{t_f M} \begin{bmatrix} I \\ Q_{1f} \\ \vdots \\ Q_{mf} \end{bmatrix}$.

**Theorem 1.** *A unique Nash equilibrium based self-enforcing control strategy that can guarantee the convoy control exists to the differential game in (11) and (13) iff there is a unique solution to the following coupled (asymmetric) Riccati differential equations:*

$$\dot{P}_i + P_i A + A^T P_i - P_i \sum_{j=1}^{n} S_j P_j + Q_i = o, \quad P_i(t_f) = Q_{if} \quad (14)$$

*Proof.* See [23], chapter 7. □

**Corollary 1.** *A unique solution to the coupled (asymmetric) Riccati differential equations (14) exists iff there is inverse for the matrix $H(t_f)$.*

*Proof.* See [23], chapter 7. □

The unique open-loop Nash equilibrium solution to the game in (11) and (13) is given by:

$$u_i = -R_{ii}^{-1} B_i^T P_i z \quad (15)$$

where
$$z = \Omega(t, 0) z(0), \quad \dot{\Omega}(t, 0) = A_{cl} \Omega(t, 0), \quad \Omega(0, 0) = I$$

and
$$A_{cl} = A - \sum_{j=1}^{m} S_j P_j(0)$$

is the closed-loop system matrix.

As a consequence of the corollary, if it can be proved that the matrix $H(t_f)$ has an inverse, then, a unique solution to the coupled Riccati differential equations (14), $P_i$ exists and therefor so does the self-enforcing convoy control strategy $u_i$. It follows from the theorem that the necessary and sufficient condition for the solvability of the coupled Riccati differential equation is matrix $H(t_f)$ being invertible.

In the next section, We reformulate this model using the relative dynamics between the vehicles. In the relative dynamics based convoy control problem, the coupling between the neighbor vehicles is in the system dynamics equation and the cost functions are uncoupled. The arisen matrix differential equations here are uncoupled, thus analytically tractable to investigate for the existence and uniqueness of a solution. We note that the presented relative dynamics system based optimal control problem carries all the features of the individual dynamic system based game problem.

## IV. Differential Game Conversion to Optimal Control

The convoy control as a differential game in (11) and (13) includes the coupling terms in each player's cost function that results in the coupling among the corresponding Riccati differential equations in (14). One can form the relative dynamics between the directly connected players (i.e., vehicles), so the coupling is relocated to the system dynamics equation, and the cost functions remain uncoupled. Consequently, the arisen matrix differential equations here will be uncoupled, thus analytically tractable to investigate the existence and uniqueness of the solution. Relative dynamics-based formation problems have been studied in the literature (see, [27]–[29]).

Based on the orders of the edges in $\mathcal{E}$ for $\forall (i,j) \in \mathcal{E}$, we define the relative dynamics between vehicle $i$ and vehicle $j$ as follows:

$$x_k = q_i - q_j - d_{ij}, \quad e_k = u_i - u_j \quad (16)$$

where $k \in \{1, \ldots, n = |\mathcal{E}|\}$.

Let $x = [x_1^T, \ldots, x_n^T, \dot{x}_1^T, \ldots, \dot{x}_n^T]^T \in \mathbb{R}^{4n}$. The complete system relative dynamics can be expressed in the following state-space form:

$$\dot{x} = (\mathcal{A} \otimes I_2)x + \sum_{i=1}^{n}(\mathcal{B}_i \otimes I_2)e_i \quad (17)$$

where $\mathcal{A} = \begin{bmatrix} \mathbf{0} & I_n \\ \mathbf{0} & \mathbf{0} \end{bmatrix} \in \mathbb{R}^{2n}$, $\mathcal{B}_i = [\mathbf{0}_n, \beta_i^T]^T \in \mathbb{R}^{2n}$ and $\beta_i = [0, \ldots, 1, \ldots 0]^T \in \mathbb{R}^n$. Here, the relative distances and relative velocities are the states, and relative accelerations are the control inputs.

The desired convoy formation can be acquired and kept if and only if the states of the relative dynamics system approach zero, i.e.,

$$\lim_{t \to \infty} x = 0 \quad (18)$$

The relative dynamics state error for the $i$th relative dynamics is defined as:

$$\mu_i(\|x_i\|^2 + \|\dot{x}_i\|^2) = x^T(I_2 \otimes \mathcal{W}_i \otimes I_2)x \quad (19)$$

where $\mathcal{W}_i = \mathrm{diag}(0, \ldots, \mu_i, \ldots, 0)$ and $\mu_i$ is the $i$th weight in the set $\{\ldots, \mu_{ij}, \ldots\}$ ordered as same as the edges in $\mathcal{E}$. Note that (19) counterparts the individual dynamics state error (12).

The finite horizon cost function for the $i$th relative dynamics is:

$$\mathcal{J}_i = x^T(t_f)(I_2 \otimes \mathcal{W}_{if} \otimes I_2)x(t_f) + \\ \int_0^{t_f}(x^T(I_2 \otimes \mathcal{W}_i \otimes I_2)x + \sum_{j=1}^{n} e_j^T(r_{ij} \otimes I_2)e_j)dt \quad (20)$$

where $\mathcal{W}_{if} = \mathrm{diag}(0, \ldots, \omega_i, \ldots, 0)$ and $\omega_i$ is the $i$th weight in the set $\{\ldots, \omega_{ij}, \ldots\}$ ordered as same as the edges in $\mathcal{E}$. Here, $r_{ii} > 0$ and for convenience, after this we denote $r_{ii}$ as $r_i$.

Applying the necessary conditions for optimality, we obtain the open-loop optimal control law for the optimal control problem (17) and (20) as:

$$e_i = \frac{1}{r_i}(\mathcal{B}_i^T \mathcal{P}_i \otimes I_2)x \quad (21)$$

Here, $\mathcal{P}_i$ is the solution to the symmetric Riccati differential equation:

$$\dot{\mathcal{P}}_i + \mathcal{P}_i(\mathcal{A} \otimes I_2) + (\mathcal{A}^T \otimes I_2)\mathcal{P}_i - \mathcal{P}_i(\mathcal{S}_i \otimes I_2)\mathcal{P}_i + \\ (I_2 \otimes \mathcal{W}_i) = \mathbf{0}, \quad \mathcal{P}_i(t_f) = I_2 \otimes \mathcal{W}_{if} \quad (22)$$

where $\mathcal{S}_i = \frac{1}{r_i}\mathcal{B}_i\mathcal{B}_i^T$.

Let $\mathcal{M} = \begin{bmatrix} -\mathcal{A} \otimes I_2 & \mathcal{S}_1 \otimes I_2 & \cdots & \mathcal{S}_n \otimes I_2 \\ \mathcal{Q}_1 \otimes I_2 & \mathcal{A}^T \otimes I_2 & \mathbf{0} & \mathbf{0} \\ \vdots & & \ddots & \mathbf{0} \\ \mathcal{Q}_m \otimes I_2 & \mathbf{0} & \mathbf{0} & \mathcal{A}^T \otimes I_2 \end{bmatrix}$ and

$\mathcal{H}(t_f) = \begin{bmatrix} I_{4n} & \mathbf{0} \end{bmatrix} e^{t_f \mathcal{M}} \begin{bmatrix} I \\ I_2 \otimes \mathcal{W}_{1f} \otimes I_2 \\ \vdots \\ I_2 \otimes \mathcal{W}_{nf} \otimes I_2 \end{bmatrix}$ be the counterparts of $M$ and $H(t_f)$ in the differential game problem, respectively.

Here, as the necessary condition to the solvability of symmetric Riccati differential equations (22), it must be checked whether or not the matrix $\mathcal{H}(t_f)$ has an inverse. The rest of this section is dedicated to find the answer.

Based on the Kronecker operator definition, we have:

$$\mathcal{M} = \mathfrak{M} \otimes I_2, \quad \mathcal{H} = \mathfrak{H} \otimes I_2 \quad (23)$$

where $\mathfrak{M} = \begin{bmatrix} -\mathcal{A} & \mathcal{S}_1 & \cdots & \mathcal{S}_n \\ \mathcal{Q}_1 & \mathcal{A}^T & \mathbf{0} & \mathbf{0} \\ \vdots & & \ddots & \mathbf{0} \\ \mathcal{Q}_m & \mathbf{0} & \mathbf{0} & \mathcal{A}^T \end{bmatrix}$ and $\mathfrak{H}(t_f) = \begin{bmatrix} I_{2n} & \mathbf{0} \end{bmatrix} e^{t_f \mathcal{M}} \begin{bmatrix} I \\ I_2 \otimes \mathcal{W}_{1f} \\ \vdots \\ I_2 \otimes \mathcal{W}_{nf} \end{bmatrix}$

From (5), matrix $\mathcal{H}(t_f)$ is invertible if and only if the matrix $\mathfrak{H}(t_f)$ is invertible. Eventually, we can guarantee the existence of convoy control as the optimal control problem (17) and (13) as well as the differential game problem (11) and (13) by proving the inverse of $\mathfrak{H}(t_f)$ exists. To that end, the following lemma is introduced first.

**Lemma 1.** *The following statements about the matrix $\mathfrak{M}$ hold:*

1) *Its eigenvalues are $\mathbf{0}_{2n(n-1)}, \ldots, \lambda_i, \overline{\lambda}_i, -\lambda_i, -\overline{\lambda}_i, \ldots$ where*

$$\lambda_i = \sqrt{\frac{\mu_i + \sqrt{\mu_i^2 - 4\mu_i r_i}}{2r_i}}, \quad i = 1, \cdots, n \quad (24)$$

*and $\overline{\lambda}_i$ denotes the conjugate of $\lambda_i$.*

2) *It is defective (i.e., non-diagonalizable).*
3) *Its Jordan canonical form is:*

$$J = I_{n(n-1)} \otimes J_2(0) \oplus \cdots \oplus J_1(\lambda_i) \oplus \cdots \quad (25)$$

*where $J_i(.)$ is the Jordan block of size $i$.*

4) *The right eigenvector and the right generalized eigenvector associated with eigenvalue zero have the form $[\mathbf{0}_{2n}, *]^T$. The left eigenvector and the left generalized eigenvector have the form $[\mathbf{0}_{2n}, *]$.*

5) Let $\boldsymbol{v}_i$ and $\boldsymbol{w}_i$ be the right and left eigenvector associated with the eigenvalue $\lambda_i$, respectively. Then,

$$\boldsymbol{v}_i = [(1, -\lambda_i) \otimes \boldsymbol{\sigma}_i^T, \boldsymbol{o}_{2n}, \ldots, \\ (\lambda_i^{-1}, \lambda_i^{-2} - 1) \otimes \boldsymbol{\sigma}_i^T \boldsymbol{\mathcal{W}}_i, \ldots, \boldsymbol{o}_{2n}]^T \quad (26)$$

$$\boldsymbol{w}_i = [(\frac{\mu_i}{r_i}\lambda_i^{-1} - \lambda_i, 1) \otimes \boldsymbol{\sigma}_i^T, \boldsymbol{o}_{2n}, \ldots, \\ (\lambda_i^{-2}, \lambda_i^{-1}) \otimes \boldsymbol{\sigma}_i^T \boldsymbol{\delta}_i, \ldots, \boldsymbol{o}_{2n}]^T \quad (27)$$

where $\boldsymbol{\sigma}_i = [0, .., \varpi_i, .., 0]^T$ and $\varpi_i$ is calculated from:

$$\varpi_i^2 = \frac{1}{2}\lambda_i^3 (\frac{\mu_i}{r_i} - \lambda_i^4)^{-1} \quad (28)$$

*Proof.* See Appendix A. □

Define

$$\boldsymbol{\Phi} = \begin{bmatrix} \boldsymbol{o}_{2n \times 2n(n-1)} & \cdots, \boldsymbol{v}_i, \cdots \\ * & \end{bmatrix}, \boldsymbol{\Psi} = \begin{bmatrix} \boldsymbol{o}_{2n(n-1) \times 2n}, * \\ \vdots \\ \boldsymbol{w}_i \\ \vdots \end{bmatrix}$$

where * are the elements/blocks to be not concerned.

Any defective matrix such as $\mathfrak{M}$ can be factored into the Jordan canonical form:

$$\mathfrak{M} = \boldsymbol{\Phi} \boldsymbol{J} \boldsymbol{\Psi} \quad (29)$$

where the generalized modal matrices $\boldsymbol{\Phi}$ and $\boldsymbol{\Psi}$ are constituted from the vectors introduced in lemma 1.4-5. We present the following remarks before continuing.

*Remark* 1.

$$[\boldsymbol{I}_{2n} \quad \boldsymbol{o}]\boldsymbol{\Phi} = [\boldsymbol{o}_{2n \times 2n(n-1)} \cdots, \begin{pmatrix} 1 \\ -\lambda_i \end{pmatrix} \otimes \boldsymbol{\sigma}_i, \cdots]$$

*Remark* 2.

$$e^{t_f \boldsymbol{J}} = \text{diag}(\boldsymbol{I}_{n(n-1)} \otimes e^{t_f \boldsymbol{J}_2(0)}, \ldots, e^{t_f \lambda_i}, \ldots)$$

*Remark* 3.

$$[\boldsymbol{I}_{2n} \quad \boldsymbol{0}]\boldsymbol{\Phi} e^{t_f \boldsymbol{J}} = [0_{2n \times 2n(n-1)} \cdots, e^{t_f \lambda_i} \begin{pmatrix} 1 \\ -\lambda_i \end{pmatrix} \otimes \boldsymbol{\sigma}_i, \cdots]$$

*Remark* 4.

$$\boldsymbol{\Psi} \begin{bmatrix} \boldsymbol{I} \\ \boldsymbol{I}_2 \otimes \boldsymbol{\mathcal{W}}_{1f} \\ \vdots \\ \boldsymbol{I}_2 \otimes \boldsymbol{\mathcal{W}}_{nf} \end{bmatrix} = \\ \begin{bmatrix} * \\ \vdots \\ (-\lambda_i + \frac{\mu_i}{r_i}\lambda_i^{-1} + \frac{w_i}{r_i}\lambda_i^{-2} \quad 1 + \frac{w_i}{r_i}\lambda_i^{-1}) \otimes \boldsymbol{\sigma}_i^T \\ \vdots \end{bmatrix}$$

Now, let

$$\boldsymbol{K}(\lambda_i, t_f) = e^{t_f \lambda_i} \begin{bmatrix} \frac{w_i}{r_i}\lambda_i^{-2} + \frac{\mu_i}{r_i}\lambda_i^{-1} - \lambda_i & 1 + \frac{w_i}{r_i}\lambda_i^{-1} \\ \lambda_i^{-2} + \frac{w_i}{r_i}\lambda_i^{-1} - \frac{\mu_i}{r_i} & -\lambda_i - \frac{w_i}{r_i} \end{bmatrix} \otimes \boldsymbol{\sigma}_i \boldsymbol{\sigma}_i^T$$

By using remarks 1-4, we can rewrite the matrix $\mathfrak{H}(t_f)$ in terms of $\boldsymbol{K}(\lambda_i, t_f)$ as:

$$\mathfrak{H}(t_f) = \begin{bmatrix} \boldsymbol{I}_{2n} & \boldsymbol{o} \end{bmatrix} \boldsymbol{\Phi} e^{t_f \boldsymbol{J}} \boldsymbol{\Psi} \begin{bmatrix} \boldsymbol{I} \\ \boldsymbol{I}_2 \otimes \boldsymbol{\mathcal{W}}_{1f} \\ \vdots \\ \boldsymbol{I}_2 \otimes \boldsymbol{\mathcal{W}}_{nf} \end{bmatrix} = \\ \sum_{i=1}^{n}(\boldsymbol{K}(\lambda_i, t_f) + \boldsymbol{K}(\overline{\lambda}_i, t_f) + \boldsymbol{K}(-\lambda_i, t_f) + \boldsymbol{K}(-\overline{\lambda}_i, t_f))$$

The following theorem reveals that $\mathfrak{H}(t_f)^{-1}$ exists and is known, therefore the optimal control problem (17) and (20) associated to $\mathfrak{H}(t_f)$ can be solved.

**Theorem 2.** : *The matrix $\mathfrak{H}(t_f)$ is real and nonsingular. The inverse of $\mathfrak{H}(t_f)$ can be obtained as:*

$$\mathfrak{H}^{-1}(t_f) = \phi \boldsymbol{\Lambda}^{-1} \psi \quad (30)$$

*where*

$$\boldsymbol{\Lambda} = \text{diag}(\cdots, \delta_i, \cdots), \boldsymbol{\phi} = [\cdots, \boldsymbol{y}_i, \cdots], \psi = \begin{bmatrix} \vdots \\ \tilde{\boldsymbol{y}}_i \\ \vdots \end{bmatrix} \quad (31)$$

$$\delta_i = \frac{1}{2}(h_i + \hat{h}_i \pm \sqrt{(h_i + \hat{h}_i)^2 - 4(h_i\hat{h}_i - \tilde{h}_i\check{h}_i)}), i = 1, \cdots, n \quad (32)$$

$$\boldsymbol{y}_i = [0, \cdots, y_i, \cdots, 0, -\frac{h_i - \delta_i}{\tilde{h}_i} y_i, \cdots, 0], \quad i = 1, \cdots, n \quad (33)$$

$$\boldsymbol{y}_i = [0, \cdots, y_i, \cdots, 0, -\frac{\check{h}_i}{\hat{h}_i - \delta_i} y_i, \cdots, 0], \quad i = n+1, \cdots, 2n \quad (34)$$

$$\tilde{\boldsymbol{y}}_i = [0, \cdots, \tilde{y}_i, \cdots, 0, -\frac{h_i - \delta_i}{\check{h}_i} \tilde{y}_i, \cdots, 0], \quad i = 1, \cdots, n \quad (35)$$

$$\tilde{\boldsymbol{y}}_i = [0, \cdots, \tilde{y}_i, \cdots, 0, -\frac{\tilde{h}_i}{\hat{h}_i - \delta_i} \tilde{y}_i, \cdots, 0], \quad i = n+1, \cdots, 2n \quad (36)$$

*Proof.* We divide the proof into three parts. First, we show that $\mathfrak{H}(t_f)$ is real. Second, we prove that it is nonsingular. Third, we obtain its inverse.

1) $\mathfrak{H}(t_f)$ is real:
As $\overline{\lambda}_i$ are $\lambda_i$ the complex conjugate of each other, we have:

$$\Re(\boldsymbol{K}(\overline{\lambda}_i, t_f)) = \Re(\boldsymbol{K}(\lambda_i, t_f)) \quad (37)$$

$$\Im(\boldsymbol{K}(\overline{\lambda}_i, t_f)) = -\Im(\boldsymbol{K}(\lambda_i, t_f)) \quad (38)$$

where $\Re(.)$ and $\Im(.)$ denote the real and imaginary parts, respectively. Therefore,

$$\boldsymbol{K}(\lambda_i, t_f) + \boldsymbol{K}(\overline{\lambda}_i, t_f) = 2\Re(\boldsymbol{K}(\lambda_i, t_f)) \quad (39)$$

and

$$\mathfrak{H}(t_f) = 2\sum_{i=1}^{n}(\Re(\boldsymbol{K}(\lambda_i, t_f)) + \Re(\boldsymbol{K}(-\lambda_i, t_f))) \quad (40)$$

2) $\mathfrak{H}(t_f)$ is nonsingular:

The eigenvalues of $\mathfrak{H}(t_f)$ are the roots of its characteristic polynomial $\rho(\delta)$:

$$\begin{aligned}\rho(\delta) &= \det(\mathfrak{H}(t_f) - \delta \boldsymbol{I}) \\ &= \det(\mathrm{diag}(\cdots,(h_i-\delta)(\hat{h}_i-\delta)-\tilde{h}_i\check{h}_i,\cdots)) \\ &= \prod_{i=1}^{n}(\delta^2 - (h_i+\hat{h})\delta + h_i\hat{h}_i - \tilde{h}_i\check{h}_i)\end{aligned} \quad (41)$$

Solving (41) for its roots, the eigenvalues of $\mathfrak{H}(t_f)$ are obtained as (32). As none of its eigenvalues is zero, $\mathfrak{H}(t_f)$ is nonsingular.

3) Obtaining $\mathfrak{H}(t_f)^{-1}$:

First, we show that the vectors $\boldsymbol{y}_i$ and $\tilde{\boldsymbol{y}}_i$ are the right and left eigenvectors associated with the eigenvalue $\delta_i$, respectively.

The right eigenvector satisfies:

$$(\mathfrak{H}(t_f) - \delta_i \boldsymbol{I})\boldsymbol{y}_i = 0 \quad (42)$$

Suppose $\boldsymbol{y}_i = [0,\cdots,y_i,\cdots,0,y_{n+i},\cdots,0]^T$. Then, (42) reduces to:

$$(h_i - \delta_i)y_i + \tilde{h}_i y_{n+i} = 0, \quad i=1,\cdots,n \quad (43)$$

$$\check{h}_i y_i + (\hat{h}_i - \delta_i)y_{n+i} = 0, \quad i=n+1,\cdots,2n \quad (44)$$

These equations have the parametric solutions:

$$(y_i, -\frac{h_i-\delta_i}{\tilde{h}_i}y_i), \quad (y_i, -\frac{\tilde{h}_i}{\hat{h}_i-\delta_i}y_i) \quad (45)$$

, respectively.

Suppose $\tilde{\boldsymbol{y}}_i = [0,\cdots,\tilde{y}_i,\cdots,0,\tilde{y}_{n+i},\cdots,0]^T$. The equivalent form of (42) for the left eigenvectors is:

$$\tilde{\boldsymbol{y}}_i(\mathfrak{H}(t_f) - \delta_i \boldsymbol{I}) = 0 \quad (46)$$

Similarly, the parametric solutions for the corresponding (reduced) equations are:

$$(\tilde{y}_i, -\frac{h_i-\delta_i}{\check{h}_i}\tilde{y}_i), \quad (\tilde{y}_i, -\frac{\tilde{h}_i}{\hat{h}_i-\delta_i}\tilde{y}_i) \quad (47)$$

, respectively.

Vectors $\boldsymbol{y}_i$ and $\tilde{\boldsymbol{y}}_i$ could be normalized so that $\tilde{\boldsymbol{y}}_i \boldsymbol{y}_i = 1$. Therefore, the parameters $y_i$ and $\tilde{y}_i$ are determined from:

$$y_i \tilde{y}_i = \frac{\tilde{h}_i \check{h}_i}{\tilde{h}_i \check{h}_i + (h_i-\delta_i)^2}, \quad i=1,\cdots,n \quad (48)$$

$$y_i \tilde{y}_i = \frac{(\hat{h}_i-\delta_i)^2}{\tilde{h}_i \check{h}_i + (\hat{h}_i-\delta_i)^2}, \quad i=n+1,\cdots,2n \quad (49)$$

As the constraint $\tilde{y}_i y_i = 1$ $(i=1,\ldots,2n)$ ensures $\boldsymbol{\phi}\boldsymbol{\psi} = \boldsymbol{I}$, the eigendecomposition for matrix $\mathfrak{H}(t_f)$ is given by:

$$\mathfrak{H}(t_f) = \boldsymbol{\phi}\boldsymbol{\Lambda}\boldsymbol{\psi} \quad (50)$$

As none of the eigenvalues of $\mathfrak{H}(t_f)$ is zero, $\boldsymbol{\Lambda}^{-1}$ exists, and consequently, $\mathfrak{H}^{-1}(t_f)$ is obtained as in (30).

The proof is complete. $\square$

The matrix $\mathfrak{H}(t_f)$ has the following structure:

$$\mathfrak{H}(t_f) = \begin{bmatrix} \mathrm{diag}(\cdots,h_i,\cdots) & \mathrm{diag}(\cdots,\tilde{h}_i,\cdots) \\ \mathrm{diag}(\cdots,\check{h}_i,\cdots) & \mathrm{diag}(\cdots,\hat{h}_i,\cdots) \end{bmatrix} \quad (51)$$

where $h_i$, $\tilde{h}_i$, $\check{h}_i$ and $\hat{h}_i$ are given in Appendix B.

## V. State-feedback Control

Practical situations demand a state-feedback controller as the external environment changes in response to players' decisions. Receding horizon control is a popular method to synthesize such a controller by the online repeated computation of the open-loop solution. Under this control scheme, the current state of the system (i.e., $\boldsymbol{x}(t)$) is considered as the initial state in the following receding horizon cost function:

$$\begin{aligned}\mathcal{J}_i^{RH} &= \boldsymbol{x}^T(t+t_f)(\boldsymbol{I}_2 \otimes \boldsymbol{W}_{if} \otimes \boldsymbol{I}_2)\boldsymbol{x}(t+t_f) + \\ &\int_t^{t+t_f}(\boldsymbol{x}^T(\boldsymbol{I}_2 \otimes \boldsymbol{W}_i \otimes \boldsymbol{I}_2)\boldsymbol{x} + \sum_{j=1}^{n}\boldsymbol{e}_j^T(r_{ij} \otimes \boldsymbol{I}_2)\boldsymbol{e}_j)d\tau\end{aligned} \quad (52)$$

The state-feedback receding horizon Nash control is defined as [21]:

$$\boldsymbol{e}_i^{RH} = -\frac{1}{r_i}(\boldsymbol{\mathcal{B}}_i^T \boldsymbol{\mathcal{P}}_i(0) \otimes \boldsymbol{I}_2)\boldsymbol{x} \quad (53)$$

In this scheme, a new solution is obtained based on the state-feedback at every time instant. So, the vehicles are not committed to the initial Nash equilibrium for the whole horizon length. This feature allows reconsidering the dynamicity of the automated highway.

The receding horizon approach to convoy control is implemented using the following algorithm:

---
**Algorithm 1:** Receding horizon Nash control

**Input:** individual vehicle states $\boldsymbol{q}_i(t)$, $\boldsymbol{q}_j(t)$ for $\forall(i,j) \in \mathcal{E}$, and $\boldsymbol{\mathcal{P}}_i(0)$ (found from (22))
**Output:** individual vehicle controls $\boldsymbol{u}_i(t)$ and $\boldsymbol{u}_j(t)$
**while** *convoy mission has not accomplished* **do**
  (1) form relative state $\boldsymbol{x}_k(t)$
  (2) use/update $\boldsymbol{\mathcal{P}}_i(0)$
  (3) calculate $\boldsymbol{e}_k^{RH}(t+\tau)$ from (53)
  (4) obtain $\boldsymbol{u}_i(t+\tau)$ and $\boldsymbol{u}_j(t+\tau)$ from $\boldsymbol{e}_k^{RH}(t)$
  (5) apply $\boldsymbol{u}_i(t+\tau)$ and $\boldsymbol{u}_j(t+\tau)$ to the $i$th and $j$th vehicle
  (6) $t \leftarrow t + \tau$
**end**

---

Notice that in algorithm 1, at step 2, $\boldsymbol{\mathcal{P}}_i(0)$ is calculated once and used at all the future times sampled at every $\tau > 0$ step. A recalculation of $\boldsymbol{\mathcal{P}}_i(0)$ is necessary if the communication graph changes at that particular time instant, i.e., if $\mathcal{E}(t) \neq \mathcal{E}(t+\tau)$.

We define

$$\boldsymbol{\mathcal{A}}_{cl} = \boldsymbol{\mathcal{A}} - \sum_{j=1}^{n}\boldsymbol{\mathcal{S}}_j \boldsymbol{\mathcal{P}}_j(0) \quad (54)$$

as the closed-loop system matrix. The receding horizon closed-loop system is given by:

$$\dot{\boldsymbol{x}}^{RH} = \boldsymbol{\mathcal{A}}_{cl}\boldsymbol{x} \quad (55)$$

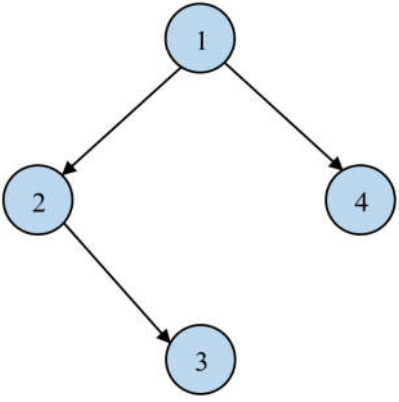

Fig. 1. Graph representation of a four-vehicle convoy.

The closed-loop system needs to be checked for stability. It is asymptotically stable if the closed-loop system matrix is Hurwitz.

**Theorem 3.** *The closed-loop system (55) is asymptotically stable.*

*Proof.* Matrix $\mathcal{A}_{cl}$ is Hurwitz if and only if for any $Q = Q^T > 0$ there is a $P = P^T > 0$ that satisfies the Lyapunov equation:
$$P\mathcal{A}_{cl} + \mathcal{A}_{cl}^T P + Q = 0 \qquad (56)$$
We select $Q = \sum_{i=1}^{n} I_2 \otimes \mathcal{W}_i$ where $Q = Q^T > 0$ holds. Let $P = \sum_{i=1}^{n} \mathcal{P}_i$. As $\mathcal{P}_i = \mathcal{P}_i^T$, we have $P^T = P$. Matrix $\mathcal{P}_i$ has a block structure similar to $\mathcal{H}(t_f)$, where only $p_i$, $\tilde{p}_i$, $\check{p}_i$ and $\hat{p}_i$ of its elements are nonzero. Therefore, $P$ is positive-definite ($P > 0$). Eventually, by substituting $\mathcal{A}_{cl}$ from (54) into (56), we obtain the algebraic version of the Riccati equation (22). This means that (56) holds for the selected matrices, and thus, the theorem proof is complete. □

## VI. SIMULATIONS

In this section, we illustrate the effectiveness of the presented convoy control model. We consider an autonomous driving scenario in a three-lane automated highway in which four vehicles are self-driving in each other's sensor detection/V2V communication range. They decide on continuing their travel as a convoy down the highway.

Suppose that the directed graph $\mathcal{G}(\mathcal{V}, \mathcal{E})$ (in Fig. 1) where $\mathcal{V} = \{1, 2, 3, 4\}$ and $\mathcal{E} = \{(1, 2), (1, 4), (2, 3)\}$ represents the convoy communication topology. Desired convoy geometric shape is defined by the offset vectors among the immediate neighbors specified in $\mathcal{E}$ as $d_{12} = [-2, -4]^T$ and $d_{14} = d_{23} = [2, -4]^T$.

Based on the orders of the edges in $\mathcal{E}$, the relationship between the relative-dynamics system and the individual vehicle dynamics system can be stated as:
$$\begin{bmatrix} x_1 \\ x_2 \\ x_3 \end{bmatrix} = \begin{bmatrix} 1 & -1 & 0 & 0 \\ 1 & 0 & 0 & -1 \\ 0 & 1 & -1 & 0 \end{bmatrix} \begin{bmatrix} q_1 \\ q_2 \\ q_3 \\ q_4 \end{bmatrix} - \begin{bmatrix} d_{12} \\ d_{14} \\ d_{23} \end{bmatrix} \qquad (57)$$

where the matrix term is $-D^T$. Likewise, a matrix equation can be formed to map the control signals from the relative dynamics system to the individual dynamics system. Numerical algorithms solve (57) and its counterpart equation for the control signals.

The initial positions and velocities of the vehicles are set to $q_1 = [1, 5]^T$, $q_2 = [1, 0]^T$, $q_3 = [5, 0]^T$, $q_4 = [3, 0]^T$, $\dot{q}_1 = [0, 2]^T$, $\dot{q}_2 = [0, 2]^T$, $\dot{q}_3 = [0, 2]^T$ and $\dot{q}_4 = [0, 2]^T$. A horizon of $t_f = 0.3$ with the update step of $\tau = 0.1$ are used. The initial longitudinal velocities are given non-zero values as the vehicles are supposed to be in the driving mode at the time of agreement on the convoy formation.

Let $\mathcal{W}_1 = \text{diag}(1, 0, 0)$, $\mathcal{W}_2 = \text{diag}(0, 1, 0)$, $\mathcal{W}_3 = \text{diag}(0, 0, 1)$, $\mathcal{W}_{if} = 5\mathcal{W}_i$ and $r_i = 1$ for $i = 1, 2, 3$ as the weighting parameters of the model. According to theorem 2, the matrix $\mathcal{H}(t_f)$ is real, nonsingular, and its inverse can be obtained from (30). Thus, a unique solution to the symmetric Riccati differential equation (22) exists. By the ODE45 solver in Matlab software, the symmetrical Riccati differential equations in (22) are solved using the terminal value and the backward iteration within the specified terminal time. Notice that the longer is the receding horizon length, the more accurate are the ODE45 solutions, and therefore, the control inputs are more precise. The finite horizon length considered here provides fairly accurate and satisfactory solutions. Finally, receding horizon control signals are calculated from (60). Based on theorem 3, the closed-loop system is asymptotically stable, i.e., the satisfaction of (18) is guaranteed, and thus, the states of the relative dynamics system approach zero.

We analyze the autonomous convoy driving scenario under three assumptions. First, we assume that the vehicles care only about the convoy acquisition and do not consider the state-space constraints. The second assumption is that they also pay attention to collision avoidance. The third assumption is that the state-space constraints of collision avoidance and driving in lane are taken into account. It is worth mentioning state-space constraints are non-convex and cannot be treated analytically within the proposed model.

### A. Convoy Acquisition

The convoy acquisition phase is planned for the time interval $t \in [0, 10]$. Fig. 2 shows the trajectories of the vehicles forming desired convoy formation over this period. The relative dynamics system time evolution is given in Fig. 3 which shows the longitudinal and lateral positions, velocities, and control inputs ($x_i$, $\dot{q}_i$, $e_i$, $i \in 1, 2, 3$) approaching zero. These results demonstrate that the relative dynamics system control framework successfully drives the vehicles to desired convoy formation. It is worth mentioning the presented control scheme easily allows lane-change maneuvers. In other words, vehicles can perform lane-change maneuvers simultaneously under the presented convoy acquisition design.

### B. Collision Avoidance

Collision avoidance (while performing a lane-change maneuver) and driving in lane are two critical state constraints that must be taken care of by ADAS and ADS when driving on automated highways. Especially, lane-change maneuver is one of the riskiest driving tasks. As it can be seen in Fig. 2, since collision avoidance was not considered when designing the convoy formation, some vehicles change their lanes without any necessity. This reckless driving increases the chance of a possible collision with other vehicles.

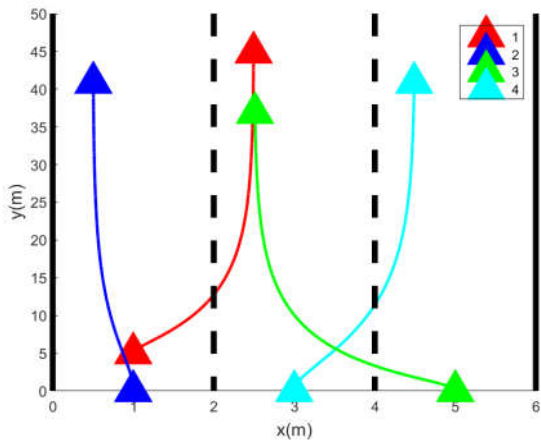

Fig. 2. Evolution of longitudinal and lateral positions of the vehicles in the acquisition phase.

The time histories of distances among all vehicle pairs for the convoy acquisition scenario in Fig. 2 are given in Fig. 4. Although there is neither a collision nor an immediate collision risk, vehicles 3 and 4 are only $1.58m$ apart at $t = 0.6s$ which can be considered a low collision risk. It is seen that for these two vehicles changing their lanes is nonnecessary. Therefore, it is reasonable to modify the convoy formation design to prevent any risk of collision. In the state-space constraints-free convoy acquisition (see Fig. 2), vehicle 3 is located at the initial position $q_3 = [5, 0]^T$, and vehicle 4 is located at the initial position $q_4 = [3, 0]^T$. Redefining $q_3$, $q_4$, as $q_3 = [3, 0]^T$, $q_4 = [5, 0]^T$, the same convoy formation is acquired while unnecessary lane changes are prevented. The evolution of convoy acquisition and the time histories of distances between all vehicle pairs under the mentioned collision avoidance consideration are demonstrated in Fig 5 and Fig 6, respectively. It can be seen that a collision risk-free desired convoy is formed.

*C. Lane-Keeping*

Lane-keeping is another vital feature of ADAS and DAS. These systems find the suitable control input to keep the vehicle in the lane. To meet this condition, we assume a virtual leader that drives on one of the lane boundaries. This virtual leader has the same dynamics as the actual leading vehicle in the convoy and only generates a reference signal for the leading vehicle to observe its lane-keeping performance by maintaining a lateral distance to it. The four-vehicle convoy information topology graph with the virtual leader (i.e., vertex 0) is shown in Fig. 7. Assume that the virtual leader keeps the lateral distance $1m$ with respect to the leading vehicle (i.e., vehicle 1) according to the desired offset vector $d_{01} = [1, -4]^T$. The evolution of longitudinal and lateral positions of the vehicles in presence of the virtual vehicle is demonstrated in Fig. 8. As it can be seen, the virtual leader steers the leading vehicle to the center of its lane as time goes on. Since vehicles 2, 3, and 4 have to maintain the convoy formation with the leading vehicle 1, they adjust their lateral positions and eventually, keep their lanes too. The time histories of relative lateral position displacement between the virtual and actual leading vehicles are shown in Fig. 9. It can be seen that the leading vehicle reaches a lateral distance of more than $0.92m$ at the end of simulation time which means it drives almost in the center of the lane. It is worth mentioning that this approach is not a leader-follower control structure as in [30] in which a (virtual or actual) leader generates a reference signal for other vehicles to follow and keep a formation.

## VII. CONCLUSION

In this paper, we addressed the autonomous convoy control problem on automated highways. The behavior of the vehicles as a convoy was modeled as a non-cooperative differential game problem. This model required solving a set of coupled Riccati matrix differential equations in which the existence and uniqueness of their solutions depend on whether or not a particular matrix has an inverse. By converting the differential game to an equivalent optimal control problem, it was shown that the counterpart matrix has an inverse, and thus optimal control law exists. Simulations on a four-vehicle convoy demonstrated the effectiveness of the models and solutions under the receding horizon control scheme. The results showed desired convoy formation was acquired, and the relative system states and control inputs approached zero. In addition, the simulations showed that the proposed convoy control allows handling state-space constraints such as collision avoidance and staying in lane.

## APPENDIX A
## PROOF OF LEMMA 1

1. By definition, the eigenvalues of a matrix are the roots of its characteristic polynomial. The characteristic polynomial for $\mathcal{M}$ can be formed as follows:

$$\rho(\lambda) = \det(\mathcal{M} - \lambda \boldsymbol{I}) \tag{58}$$

where $\lambda$ is an unknown variable representing the unknown eigenvalues. Let,

$$\boldsymbol{D} = \begin{bmatrix} \mathcal{A}^T & \mathbf{0} & \mathbf{0} \\ \mathbf{0} & \ddots & \mathbf{0} \\ \mathbf{0} & \mathbf{0} & \mathcal{A}^T \end{bmatrix}, \boldsymbol{S} = \begin{bmatrix} \mathcal{S}_1 & \cdots & \mathcal{S}_n \end{bmatrix},$$

$$\boldsymbol{W} = \begin{bmatrix} \boldsymbol{I}_2 \otimes \mathcal{W}_1 \\ \vdots \\ \boldsymbol{I}_2 \otimes \mathcal{W}_n \end{bmatrix}$$

The determinant of the block matrix $\mathcal{M} - \lambda \boldsymbol{I}$ can be reduced as [32]:

$$\det(\mathcal{M} - \lambda \boldsymbol{I}) = \det \begin{bmatrix} -\mathcal{A} - \lambda \boldsymbol{I} & \boldsymbol{S} \\ \boldsymbol{W} & \boldsymbol{D} - \lambda \boldsymbol{I} \end{bmatrix} = \\ \det(-\mathcal{A} - \lambda \boldsymbol{I}) \det(\boldsymbol{D} - \lambda \boldsymbol{I} - \boldsymbol{W}(-\mathcal{A} - \lambda \boldsymbol{I})^{-1} \boldsymbol{S}) \tag{59}$$

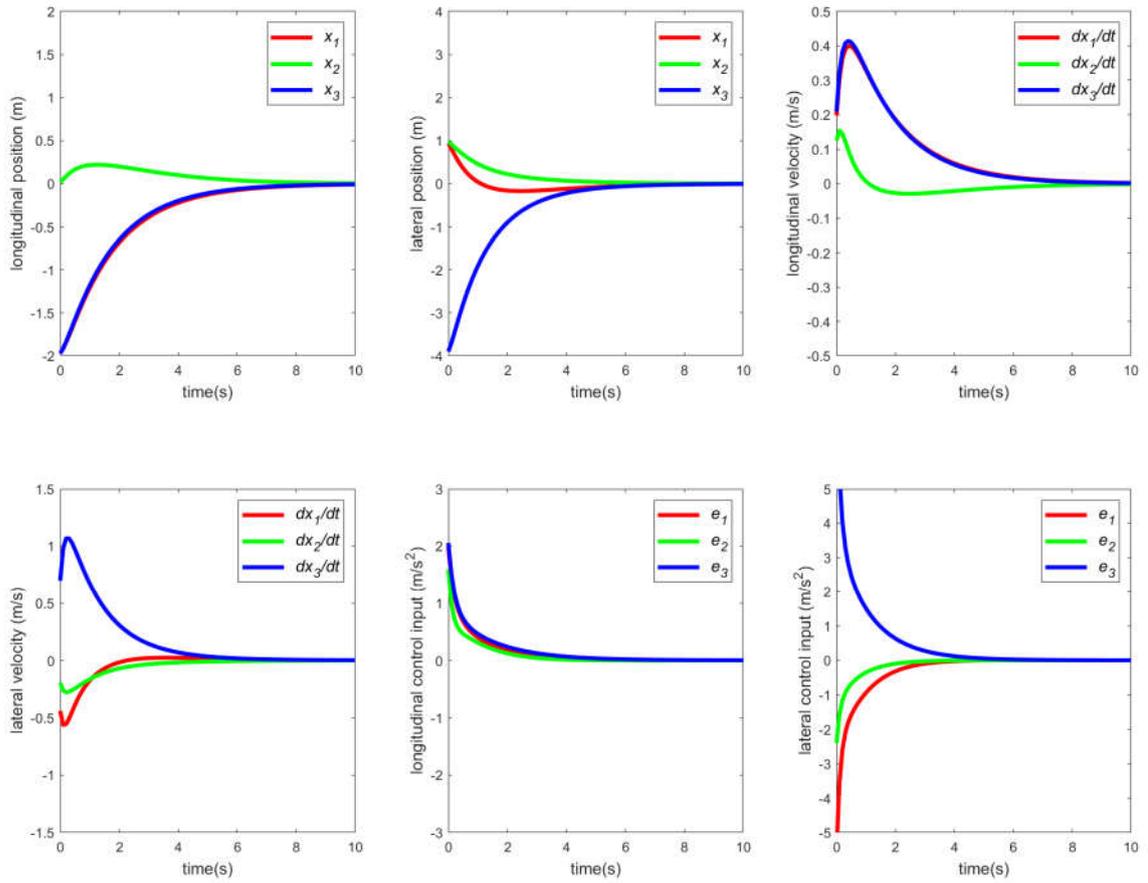

Fig. 3. Longitudinal and lateral time histories of relative dynamics system.

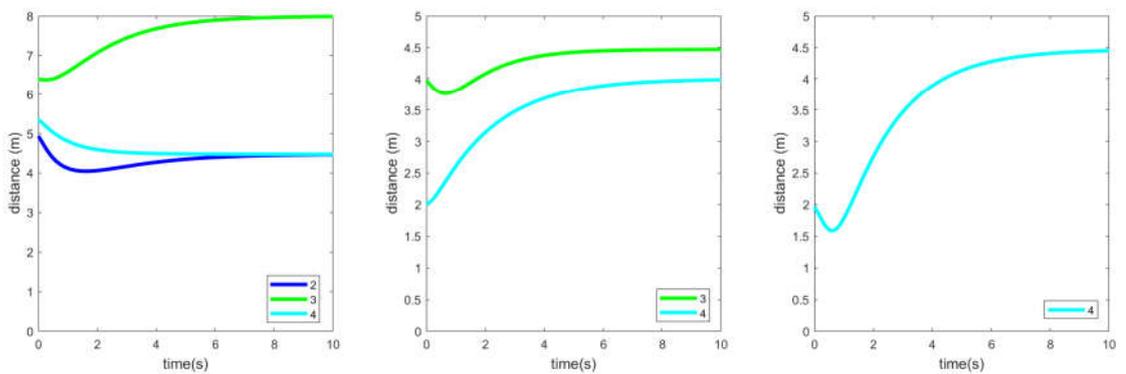

Fig. 4. Time histories of distances among the vehicles when the state-space constraints are not considered. The left, middle and right figures show the distances of $i$) vehicle 1 to vehicles 2,3, and 4, $ii$) vehicle 2 to vehicles 3 and 4, $iii$) vehicle 3 to 4.

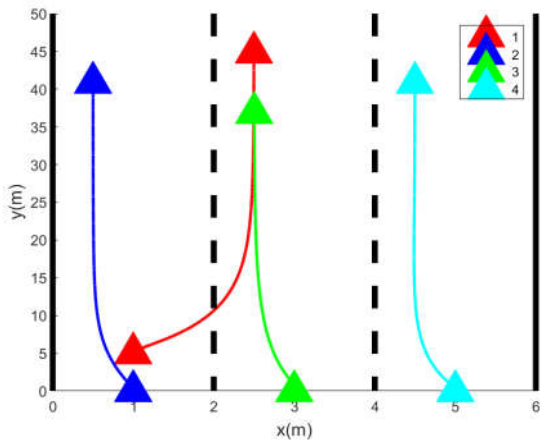

Fig. 5. Evolution of longitudinal and lateral positions of the vehicles under the collision avoidance considered design.

Based on the Kronecker property (7), we have:

$$\det(\boldsymbol{\mathcal{A}} - \lambda \boldsymbol{I}_{2n}) = \det(-\begin{bmatrix} 0 & 1 \\ 0 & 0 \end{bmatrix} \otimes \boldsymbol{I}_n - \lambda \boldsymbol{I}_2 \otimes \boldsymbol{I}_n) = \\ \det(-\begin{bmatrix} 0 & 1 \\ 0 & 0 \end{bmatrix} \otimes \boldsymbol{I}_n) = (\det(-\begin{bmatrix} 0 & 1 \\ 0 & 0 \end{bmatrix}))^n = \lambda^{2n} \quad (60)$$

The inverse of the block matrix $-\boldsymbol{\mathcal{A}} - \lambda \boldsymbol{I}$ can be obtained as [33]:

$$(-\boldsymbol{\mathcal{A}} - \lambda \boldsymbol{I}_{2n})^{-1} = (-\begin{bmatrix} \lambda & 1 \\ 0 & \lambda \end{bmatrix} \otimes \boldsymbol{I}_n)^{-1} = \\ -\begin{bmatrix} \lambda & 1 \\ 0 & \lambda \end{bmatrix}^{-1} \otimes \boldsymbol{I}_n = \begin{bmatrix} -\lambda^{-1} & \lambda^{-2} \\ 0 & -\lambda^{-1} \end{bmatrix} \otimes \boldsymbol{I}_n \quad (61)$$

We notice

$$\boldsymbol{\mathcal{S}}_i = \frac{1}{r_i} \boldsymbol{\mathcal{B}}_i \boldsymbol{\mathcal{B}}_i^T = \begin{bmatrix} 0 & 0 \\ 0 & 1 \end{bmatrix} \otimes \boldsymbol{\delta}_i \quad (62)$$

where $\boldsymbol{\delta}_i = \mathrm{diag}(0, \cdots, \frac{1}{r_i}, \cdots, 0) \in \mathbb{R}^n$. We notice that $\boldsymbol{\mathcal{W}}_i \boldsymbol{\delta}_i = \mathrm{diag}(0, \cdots, \frac{\mu_i}{r_i}, \cdots, 0) \in \mathbb{R}^n$ and $\boldsymbol{\mathcal{W}}_i \boldsymbol{\delta}_j = 0$ for $i \neq j$. Using (61), we get:

$$\boldsymbol{\mathcal{W}}(-\boldsymbol{\mathcal{A}} - \lambda \boldsymbol{I})^{-1} \boldsymbol{S} = \mathrm{diag}(\cdots, \begin{bmatrix} 0 & \lambda^{-2} \\ 0 & -\lambda^{-1} \end{bmatrix} \otimes \boldsymbol{\mathcal{W}}_i \boldsymbol{\delta}_i, \cdots) \quad (63)$$

On the other hand:

$$\boldsymbol{D} - \lambda \boldsymbol{I} = \mathrm{diag}(\cdots, \begin{bmatrix} -\lambda & 0 \\ 0 & -\lambda \end{bmatrix} \otimes \boldsymbol{I}_n, \cdots) \quad (64)$$

Substituting (60), (63) and (64) in (59) yields:

$$\begin{aligned}\rho(\lambda) &= \lambda^{2n} \det(\mathrm{diag}(\cdots, \begin{bmatrix} -\lambda \boldsymbol{I}_n & -\lambda^{-2} \boldsymbol{\mathcal{W}}_i \boldsymbol{\delta}_i \\ \boldsymbol{I}_n & -\lambda \boldsymbol{I}_n + \lambda^{-1} \boldsymbol{\mathcal{W}}_i \boldsymbol{\delta}_i \end{bmatrix}, \cdots)) \\ &= \lambda^{2n} \prod_{i=1}^{n} \det \begin{bmatrix} -\lambda \boldsymbol{I}_n & -\lambda^{-2} \boldsymbol{\mathcal{W}}_i \boldsymbol{\delta}_i \\ \boldsymbol{I}_n & -\lambda \boldsymbol{I}_n + \lambda^{-1} \boldsymbol{\mathcal{W}}_i \boldsymbol{\delta}_i \end{bmatrix} \\ &= \lambda^{2n} \prod_{i=1}^{n} \det(-\lambda \boldsymbol{I}_n) \prod_{i=1}^{n} \det((\lambda^{-1} - \lambda^{-3}) \boldsymbol{\mathcal{W}}_i \boldsymbol{\delta}_i - \lambda \boldsymbol{I}_n) \\ &= \lambda^{n(n+2)} \prod_{i=1}^{n} \det(-\lambda \boldsymbol{I}_{n-1}) \prod_{i=1}^{n}((\lambda^{-1} - \lambda^{-3}) \frac{\mu_i}{r_i} - \lambda) \\ &= \lambda^{n(2n+1)} \prod_{i=1}^{n} \frac{1}{\lambda^3}(\frac{\mu_i}{r_i}(\lambda^2 - 1) - \lambda^4) \\ &= \lambda^{2n(n-1)} \prod_{i=1}^{n}(\lambda^4 - \frac{\mu_i}{r_i}\lambda^2 + \frac{\mu_i}{r_i}) \end{aligned} \quad (65)$$

The quartic polynomial in (65) has four roots given by (24), two opposites in sign, and two conjugates.

2. Each eigenvalue has a geometric multiplicity that is the number of linearly independent eigenvectors associated with it. If one of the eigenvalues has geometric multiplicity less than the algebraic multiplicity the matrix is called defective [34]. Defective matrices are non-diagonalizable.

Let $\mathcal{N}(\boldsymbol{\mathcal{M}} - \lambda \boldsymbol{I})$ denote the dimension of the null space of $\boldsymbol{\mathcal{M}} - \lambda \boldsymbol{I}$. By definition, the geometric multiplicity of the eigenvalue $\lambda$ of $\boldsymbol{\mathcal{M}}$ is the $\dim \mathcal{N}(\boldsymbol{\mathcal{M}} - \lambda \boldsymbol{I})$ that is the number of free variables in the equation:

$$(\boldsymbol{\mathcal{M}} - \lambda \boldsymbol{I}) \boldsymbol{v} = 0 \quad (66)$$

where $\boldsymbol{v}$ is a non-zero vector.

Consider the block form of the matrix $\boldsymbol{\mathcal{M}}$. Let $\boldsymbol{v} = [(\boldsymbol{\varrho}_0, \boldsymbol{\vartheta}_0)^T, (\boldsymbol{\varrho}_1, \boldsymbol{\vartheta}_1)^T, \ldots, (\boldsymbol{\varrho}_n, \boldsymbol{\vartheta}_n)^T]^T$, $\boldsymbol{\varrho}_i = [\varrho_1^i, \ldots, \varrho_n^i]^T \in \mathbb{R}^n$, $\boldsymbol{\vartheta}_i = [\vartheta_1^i, \ldots, \vartheta_n^i]^T \in \mathbb{R}^n$. For $\lambda = 0$, (66) reduces to:

$$\boldsymbol{\vartheta}_0 = \boldsymbol{\varrho}_i = 0, \quad i = 0, \cdots, n, \quad \sum_{i=1}^{n} \boldsymbol{\delta}_i \boldsymbol{\vartheta}_i = \begin{bmatrix} r_1^{-1} \varrho_1^1 \\ \vdots \\ r_n^{-1} \varrho_n^n \end{bmatrix} = 0 \quad (67)$$

Clearly, $\vartheta_i^i$ in $\boldsymbol{\vartheta}_i$ must be zero, and then, there are $n - 1$ free variables left in $\boldsymbol{\vartheta}_i$. Therefore, the total number of free variables are $n(n-1)$ and consequently:

$$\dim \mathcal{N}(\boldsymbol{\mathcal{M}}) = n(n-1) \quad (68)$$

As the geometric multiplicity of the eigenvalue zero is less than its algebraic multiplicity, $\boldsymbol{\mathcal{M}}$ is defective.

3. By definition, the integer $\zeta$ is the first integer for which:

$$\dim \mathcal{N}(\boldsymbol{\mathcal{M}} - \lambda \boldsymbol{I})^{\zeta} \quad (69)$$

stabilize. Then, $\zeta$ is the size of the corresponding Jordan block to the eigenvalue $\lambda$.

Here, $\dim \mathcal{N}(\boldsymbol{\mathcal{M}} - \lambda \boldsymbol{I})^j$ is equal to the number of free variables in the following equation:

$$(\boldsymbol{\mathcal{M}} - \lambda \boldsymbol{I})^j \boldsymbol{v} = 0 \quad (70)$$

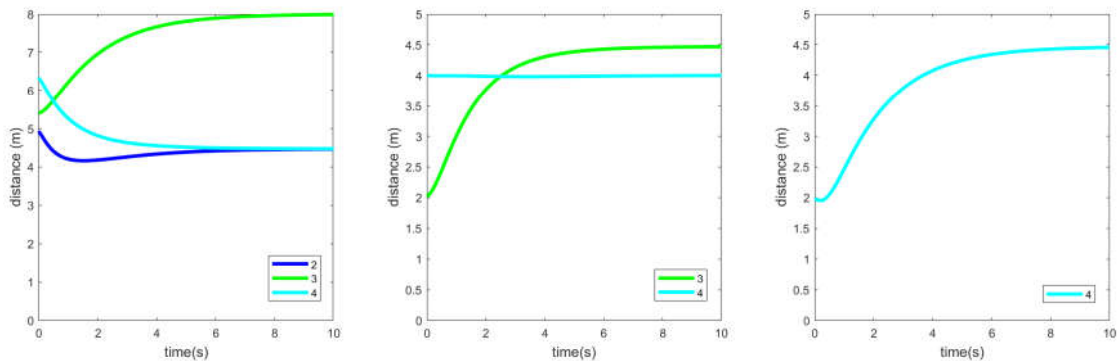

Fig. 6. Time histories of distances among the vehicles while the collision avoidance constraint is considered. The left, middle and right figures show the distances of $i$) vehicle 1 to vehicles 2,3, and 4, $ii$) vehicle 2 to vehicles 3 and 4, $iii$) vehicle 3 to 4.

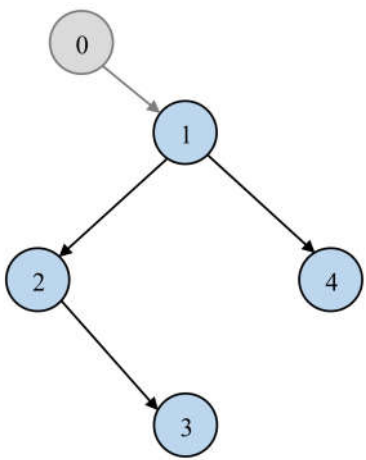

Fig. 7. Graph representation of a four-vehicle convoy with a virtual leader (i.e., vertex 0).

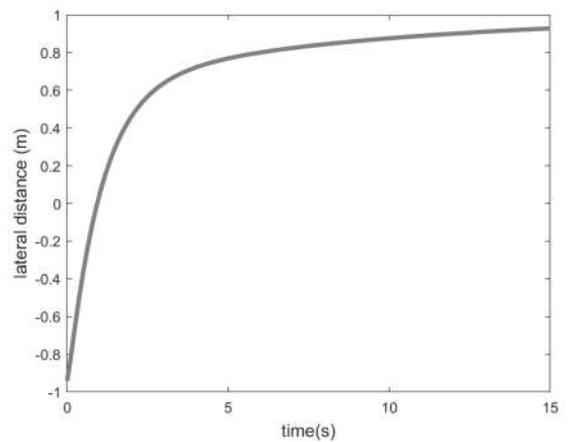

Fig. 9. Time histories of relative lateral position displacement between the virtual and the actual leading vehicle.

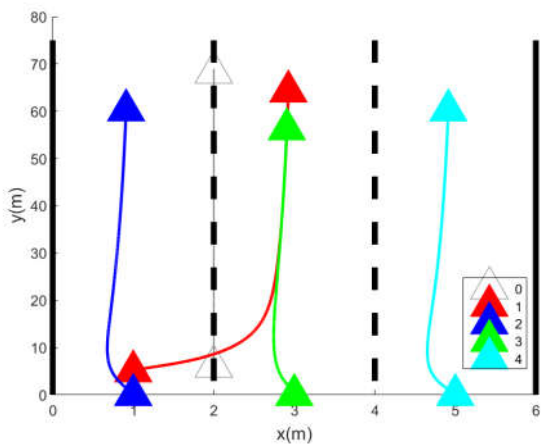

Fig. 8. Evolution of longitudinal and lateral positions of the vehicles in presence of the virtual leader.

For $\lambda = 0$, we have, $\dim \mathcal{N}(\mathcal{M})^1 = n(n-1)$. It can be verified that vector $v$ with $\varrho_0 = \vartheta_0 = 0$ and $\varrho_i = \vartheta_i = [\varrho_1^i, ..., [0]_i, ..., \varrho_n^i]^T, i = 1, ..., n$ satisfies (A14) for $j = 2$ and $j = 3$. Therefore,

$$\dim \mathcal{N}(\mathcal{M})^2 = \dim \mathcal{N}(\mathcal{M})^3 = 2n(n-1) \quad (71)$$

It is seen that $\zeta = 2$ and thus, there are $n(n-1)$ Jordan blocks of size 2.

For $\lambda \neq 0$, the number of free variables in the equation (A14) for $j = 1$ and $j = 2$ are both equal to $n$. Therefore, $\zeta = 1$ and the corresponding Jordan block is of size 1.

4. The Jordan block of size 2 (or $\zeta = 2$) means that we must find a chain of generalized eigenvectors of rank 2 for the zero eigenvalues. Notice that, by definition, a generalized eigenvector of rank 1 is an ordinary eigenvector.

An ordinary right eigenvector $v_1$ and generalized right eigenvector $v_2$ of rank 2 corresponding to eigenvalue $\lambda$ are the non-zero vectors that satisfy the following equations:

$$(\mathcal{M} - \lambda \boldsymbol{I})\boldsymbol{v}_1 = 0, \quad (\mathcal{M} - \lambda \boldsymbol{I})\boldsymbol{v}_2 = \boldsymbol{v}_1 \quad (72)$$

For $\lambda = 0$, it can be verified that:

$$\begin{aligned}\boldsymbol{v}_1 &= [(\mathbf{o}_n, \mathbf{o}_n)^T, (\mathbf{o}, \boldsymbol{\vartheta}_1)^T, \ldots, (\mathbf{o}, \boldsymbol{\vartheta}_n)^T]^T \\ \boldsymbol{v}_2 &= [(\mathbf{o}_n, \mathbf{o}_n)^T, (\boldsymbol{\vartheta}, \mathbf{o})^T, \ldots, (\boldsymbol{\vartheta}_n, \mathbf{o})^T]^T\end{aligned} \quad (73)$$

where $\boldsymbol{\vartheta}_i = [\vartheta_1^i, \cdots, [0]_i, \cdots, \vartheta_n^i]^T \in \mathbb{R}^n$.

The equivalent form of (72) for the left eigenvectors is:

$$\boldsymbol{v}_3^T(\boldsymbol{\mathcal{M}} - \lambda \boldsymbol{I}) = 0, \quad \boldsymbol{v}_4^T(\boldsymbol{\mathcal{M}} - \lambda \boldsymbol{I}) = \boldsymbol{v}_1^T \quad (74)$$

Once again, for $\lambda = 0$, it can be verified that $\boldsymbol{v}_4^T = \boldsymbol{v}_1$ and $\boldsymbol{v}_3^T = \boldsymbol{v}_2$.

5. Since for the non-zero eigenvalue $\zeta = 1$, we must find an eigenvector of rank 1. Any right eigenvector $\boldsymbol{v}_i$ must satisfy (66). For (26), we have:

$$(\boldsymbol{\mathcal{M}} - \lambda_i \boldsymbol{I}) \times$$

$$[(1, -\lambda_i) \otimes \boldsymbol{\sigma}_i^T, \mathbf{o}_{2n}, \cdots, (\frac{1}{\lambda_i}, \frac{1-\lambda_i^2}{\lambda_i^2}) \otimes \boldsymbol{\sigma}_i^T \boldsymbol{\mathcal{W}}_i, \cdots, \mathbf{o}_{2n}]^T$$

$$= [(0, \lambda_i^2 + \frac{1-\lambda_i^2}{\lambda_i^2}\frac{\mu_i}{r_i}) \otimes \boldsymbol{\sigma}_i^T, (0,0) \otimes \boldsymbol{\sigma}_i^T, \cdots, (0,0) \otimes \boldsymbol{\sigma}_i^T]^T$$

(75)

where $\sigma_i = [0, \cdots, \varpi_i, \cdots, 0]^T \in \mathbb{R}^n$. Here, the expression:

$$\lambda_i^2 + \frac{1-\lambda_i^2}{\lambda_i^2}\frac{\mu_i}{r_i} \quad (76)$$

appears to be the quartic polynomial in (65) that $\lambda_i$ is a zero of it. Therefore, (26) satisfies (66).

Any left eigenvector $\boldsymbol{w}_i$ must satisfy the first equation in (A17). It can be checked that (27) satisfies:

$$[(\frac{\mu_i}{r_i}\frac{1}{\lambda_i} - \lambda_i, 1) \otimes \boldsymbol{\sigma}_i^T, \mathbf{o}_{2n}, \cdots, (\frac{1}{\lambda_i^2}, \frac{1}{\lambda_i}) \otimes \boldsymbol{\sigma}_i^T \boldsymbol{\delta}_i, \cdots, \mathbf{o}_{2n}]^T$$

$$\times (\boldsymbol{\mathcal{M}} - \lambda \boldsymbol{I}) = 0$$

(77)

By choosing $\varpi_i$ as (28), vectors $\boldsymbol{v}_i$ and $\boldsymbol{w}_i$ could be normalized so that $\boldsymbol{w}_i \boldsymbol{v}_i = 1$.

## APPENDIX B
## ELEMENTS OF $\mathfrak{H}(t_f)$

$$h_i = \Re\{(\frac{\mu_i}{r_i} - \lambda_i^4)^{-1}(e^{t_f \lambda_i}(\frac{\omega_i}{r_i}\lambda_i + \frac{\mu_i}{r_i}\lambda_i^2 - \lambda_i^4) - e^{-t_f \lambda_i}(\frac{\omega_i}{r_i}\lambda_i - \frac{\mu_i}{r_i}\lambda_i^2 + \lambda_i^4))\}$$

$$\tilde{h}_i = \Re\{(\frac{\mu_i}{r_i} - \lambda_i^4)^{-1}(e^{t_f \lambda_i}(\lambda_i^3 + \frac{\omega_i}{r_i}\lambda_i^2) - e^{-t_f \lambda_i}(\lambda_i^3 - \frac{\omega_i}{r_i}\lambda_i^2))\}$$

$$\check{h}_i = \Re\{(\frac{\mu_i}{r_i} - \lambda_i^4)^{-1}(e^{t_f \lambda_i}(-\frac{\omega_i}{r_i}\lambda_i^2 - \frac{\mu_i}{r_i}\lambda_i^3 + \lambda_i^5) - e^{-t_f \lambda_i}(\frac{\omega_i}{r_i}\lambda_i^2 - \frac{\mu_i}{r_i}\lambda_i^3 + \lambda_i^5))\}$$

$$\hat{h}_i = \Re\{(\frac{\mu_i}{r_i} - \lambda_i^4)^{-1}(e^{t_f \lambda_i}(-\frac{\omega_i}{r_i}\lambda_i^3 - \lambda_i^4) - e^{-t_f \lambda_i}(-\frac{\omega_i}{r_i}\lambda_i^3 + \lambda_i^4))\}$$

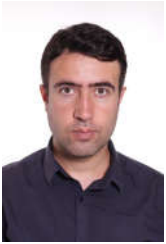

**Hossein B. Jond** received a Ph.D. degree in computer engineering from KTU, Trabzon, Turkey, in 2019. He was a post-doctoral researcher from 2019 to 2021 with the Department of Computer Science at the VSB-TUO, Ostrava, Czechia, where he is currently an Assistant Professor. His research interests include game theory and dynamic optimization. Dr. Jond was a recipient of the Turkish Government Success Scholarship in 2015, and the Best Student Paper Award in 21th International Student Conference on Electrical Engineering held in Prague in 2017.

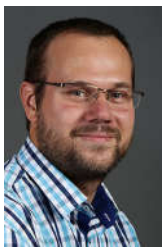

**Jan Platos** received a Ph.D. degree in Computer science in 2010. Since 2017, he has been the Head of the Department of Computer Science, Faculty of Electrical Engineering and Computer Science, VSB-TUO. He achieved the position of Full professor in 2021 at the Department of Computer Science. He has co-authored more than 200 scientific articles published in proceedings and journals. His citation report consists of 542 citations and H-index of 11 on the Web of Science, 977 citations and H-index of 15 on Scopus, and 1505 citations and H-index of 19 on Google Scholar. His primary fields of interest are machine learning, artificial intelligence, industrial data processing, text processing, data compression, bioinspired algorithms, information retrieval, data mining, data structures, and data prediction.